# Predictive Modeling of Physical and Mechanical Properties of Pervious Concrete using XGBoost

Ismail B. Mustapha[1], Zainab Abdulkareem[2], Muyideen Abdulkareem[3*], Abideen Ganiyu[4]

[1] Computer Science Department, School of Computing, Universiti Teknologi Malaysia, 81310 Johor, Malaysia
[2] Computer and Information Sciences Department, University of Strathclyde, Glasgow, G1 1XQ, United Kingdom
[3] Faculty of Engineering and Built Environment, UCSI University, 56000 Kuala Lumpur, Malaysia
[4] Department of Civil Engineering & Quantity Surveying, Military Technological College, Muscat, Oman

*E-mail: abdulkareem@ucsiuniversity.edu.my

**Abstract**. High permeability of pervious concrete (PC) makes it a special type of concrete utilised for certain applications. However, the complexity of the behaviour and properties of PC leads to costly, time consuming and energy demanding experimental works to accurately determine the mechanical and physical properties of PC. This study presents a predictive model to predict the mechanical and physical properties of PC using Extreme Gradient Boost (XGBoost). The compressive strength, tensile strength, density and porosity of PC was predicted using four models evaluated using different statistical parameters. These statistical measures are the root mean squared error (RMSE), square of correlation coefficient (R2), mean absolute error (MAE) and mean absolute percentage error (MAPE). The estimation of these properties by the XGBoost models were in agreement with the experimental measurements. The performance of XGBoost is further validated by comparing its estimations to those obtained from four corresponding support vector regression (SVR) models. The comparison showed that XGBoost generally outperformed SVR with lower RMSE of 0.58, 0.17, 0.98 and 34.97 compared to 0.74, 0.21, 1.28 and 44.06 in SVR for compressive strength, tensile strength, porosity, and density estimation respectively. Due to high correlation between the predicted and experimentally obtained properties, the XGBoost models are able to provide quick and reliable information on the properties of PC which are experimentally costly and time consuming. A feature importance and contribution analysis of the input/predictor variables showed that the cement proportion is the most important and contributory factor in the PC properties estimated.

## 1. Introduction

Pervious concrete (PC) is a specially-made concrete that contains water, Portland cement and coarse aggregate that has narrow particle size distribution. Since no fine aggregate is included, the cement paste forms a coating around the coarse aggregate, thus, a substantial part of the concrete is made up of interconnected voids. Due to this voids, pervious concrete has high permeability property that prevents stagnation of water makes it suitable in several engineering applications. Such applications include parking areas and pedestrian walkways where it allows seepage of runoffs into the ground, as well as roadways safety where it prevents skidding of vehicles [1]. Similarly, pervious concrete is used in stormwater management for purification as it acts as natural filters to reduce the amount of pollutants entering ponds, streams, lakes and rivers [2].

The application of pervious concrete cannot be overplayed in our everyday lives. However, these applications are hindered by the important porosity property that decreases the compressive strength of PC. The porosity of concrete is inversely proportional to the compressive strength. Due to the importance of compressive strength in concrete, researchers have carried out several studies to improve the compressive strength of pervious concrete and to understand other properties. For example, Yang and Jiang [3] carried out early experiments on PC pavement. The pavement materials included small-sized aggregate, superplasticiser and silica fume, and were able to produce PC with compressive strength of 50 MPa. Chen et al. [4] succeeded in producing high strength PC with 32 – 46 MPa at 28 days. The mix consisted of silica fume (SF), superplasticisers (SP), supplementary cementitious materials and polymer modification. The authors also showed that the flexural strength of PC is more sensitive to porosity than the compressive strength. Chang et al. [5] showed that PC made from alkali-activated slag cement and electric arc furnace slag had higher compressive strength than those produced by Portland cement. In addition, the sound absorption ratio attained

---

0.94 for low frequency noise. Ibrahim et al. [6] produced PC with palm oil clinkers that had slightly lowered compressive strength than the control made of coarse aggregate. Lopez-Carrasquillo and Hwang [7] improved the mechanical and physical properties of PC by using nanomaterials and fly ash. Recently, Li et al. [8] incorporated waste glass powder to increase the freeze-thaw resistance of PC. In this study, the compressive strength of the PC decreased at the early stage (28 days) and slowly increased towards the 58 days. However, the permeability decreased as the waste glass powder content increased. Yang et al. [9] produced and analysed PC made with recycled aggregate. The recycled aggregate was coated with cement paste to reinforce it due to presence of micro-cracks and weak interface transition zones. The PC had sufficient permeability and attained a compressive strength of 18.5 MPa.

In addition to improving the compressive strength, researchers have also worked on improving the porosity property of PC. Yeih et al. [10] applied electric arc furnace slag (EAFS) as aggregate in PC. Their study showed that PC made from EAFS aggregate had greater permeability coefficient and better mechanical strength than those from river gravels. Nguyen et al. [11] replaced 60% mass of aggregate with seashells. The permeability of PC remained acceptable even after a blend of silty clay and sand were applied as clogging agent. The chemical characteristics of the seashells had more effect on durability than the mechanical and physical properties. Huang et al. [12] examined the pore size distribution and permeability reduction capacity of PC.

The above studies show the complexity of PC properties and the non-linearity in its behaviour. More so, the production and design of PC is costly as well as time and energy consuming. Due to this, it is vital to apply accurate theoretical estimation of the engineering properties (wet and hardened states) of PC so as to minimize time, cost and energy wastage during the testing period. Owing to this, researchers have adopted the idea of machine learning such as artificial neural network (ANN), ANFIS, SVM and genetic algorithm to predict the properties of PC (and other types of concrete) in both wet and hardened states. For example, Adewumi et al. [13] applied four different SVR models to estimate the several properties of PC. The models were characterised with high values of correlation coefficients, low mean absolute errors and low root mean square roots. Alam et al. [14] proposed a Bayesian optimisation algorithm to estimate the shear capacity of FRP concrete members. The study applied 216 data instances and results showed the proposed method outperformed genetic algorithm [15] and ANN [16]. Imam et al. [17] estimated the compressive strength of quaternary blend concrete by using a neural network. The concrete contained Portland cement, metakaolin, fly ash and rice husk ash. The proportion of the constituents and curing days were applied as the input variables, while the output target was the compressive strength. By using three different optimization methods (Bayesian regularized, scaled conjugate gradient and Levenberg-Marquardt algorithm), the authors concluded that the Bayesian regularization function provided the best results. Salami et al. [18] proposed SVM for prediction of compressive strength of ternary blend concrete. The proportions of blast furnace slag, fly ash, superplasticiser, Portland cement, water and aggregates were used as the input variables while compressive strength was the output. By using the experimental data obtained from Yeh [19], Silva et al. [20] applied random forest, ANN and SVM models to predict the compressive strength of concrete. The predicted strength was the output parameter while the proportions of constituents were the inputs.

In this study, Extreme Gradient Boost (XGBoost) is applied to predict the properties of pervious concrete. Being an effective scalable variant of Gradient Boosting machine (GBM) that is easy to use and with very high accuracy [21]. Four different XGBoost models are developed, each to obtain the four estimated properties (density, compressive strength, tensile strength and porosity) of PC. The input variables in each XGBoost model are proportions of Portland cement and coarse aggregate, water-cement ratio and coarse aggregate size, while the properties estimated as the output are compressive strength, tensile strength, density and porosity. The predicted results by XGBoost showed high accuracy with the experimental results. In order to evaluate the performance of the XGBoost models, prediction results from SVR are also obtained as baselines. The superiority of XGBoost is shown as the predicted results were more accurate than those obtained through SVM.

## 2. Computational Methods
### 2.1. Extreme gradient boost (XGBoost)

Extreme Gradient Boosting (XGBoost) algorithm is based on the idea of "boosting", which combines all the predictions of a set of "weak" learners for developing a "strong" learner through additive training strategies [22]. XGBoost aims to prevent over-fitting as well as also optimize the computation resources. This is obtained by simplifying the objective functions that allow combining predictive and regularization terms while maintaining an optimal computational speed. Also, XGBoost has been designed to be parallelizable, computationally efficient and highly robust to overfitting. Given a dataset, the XGBoost algorithm ensembles $k \in K$ weakly learnt Classification and Regression Trees (CART). The prediction output of each weakly learnt CART is combined using additive function to obtain a strong learner as given in Equation (1):

$$\hat{y}_m = \sum_{k=1}^{K} f_k(x_m), f_k \in F, \tag{1}$$

where $f_k$ is a tree structure from the space of all possible CARTs $F$. XGBoost is iteratively learnt such that each succeeding learner tries to rectify the weakness in the prediction of its predecessor. Thus, $\hat{y}$ for time step $t$ is given as follows:

$$\hat{y}_m^{(t)} = \sum_{k=1}^{K} f_k(x_m) = \hat{y}_m^{(t-1)} + f_t(x_m), \tag{2}$$

where $f_t(x_m)$, $\hat{y}_m^{(t)}$ and $\hat{y}_m^{(t-1)}$ represent the learner additive function for time step $t$, the prediction out for time step $t$ and $t-1$ respectively. The regularized optimization objective is as in Equation (3)

$$Obj = \sum_{m}^{n} l\left(y_m, \hat{y}_m^{(t-1)} + f_t(x_m)\right) + \Omega(f_t) \tag{3}$$

where $l(.)$ is the loss function and $\Omega$ is the regularization term given as;

$$\Omega(f) = \gamma T + \frac{1}{2}\lambda \sum_{j=1}^{T} \omega^2, \tag{4}$$

where $\gamma$ and $\lambda$ are constants for controlling the degree of regularization. $T$ and $\omega$ on the other hand respectively stand for the number of leaves and the score on each leaf.

## 2.2. Support vector regression (SVR)

Since its introduction by Cortes and Vapnik in 1995 [23], support vector machine (SVM) has risen to become one of the most powerful machine learning algorithms with widespread adoption in numerous application domains. Initially introduced for classification problems, SVM has since been adapted to regression problems and called support vector regression (SVR). SVR leverages ε-insensitive loss function for the penalization of data when they are greater than ε [24]. SVR essentially learns a nonlinear mapping function to project low dimensional input-data to a high dimensional feature space [25].

Given a data $\{(x_m, y_m)\}_{m=1}^{n}$, where $x_m \in \mathbb{R}^d$ and $y_m \in \mathbb{R}$ are respectively the *m-th* input and it corresponding real-valued output. $n$ is the sample size. The SVR model output for each input data is given as:

$$g(x_m) = w^T \varphi(x_m) + b \tag{5}$$

where $\varphi$ is a nonlinear mapper of input $x_m$ to high dimensional feature space, $w$ is the weight coefficient and $b$ is the bias term. Flatness of Equation (1) is ensured by minimizing the norm of $w$ as given in Equation (6).

$$\min \frac{1}{2}\|w\|^2,$$
$$s.t \begin{cases} y_m - w^T \varphi(x_m) - b \leq \varepsilon, \\ y_m - w^T \varphi(x_m) - b \geq \varepsilon, \end{cases} \tag{6}$$

Introducing $\xi$ and $\xi^*$, two slack variables which respectively stand for the upper and lower deviations to penalise the $\varepsilon$-insensitive band give Equation (7)

$$\min \frac{1}{2}||w||^2 + C \sum_{m=1}^{n} \left( \xi_m + \xi_m^* \right),$$

$$s.t \begin{cases} y_m - w^T \varphi(x_m) - b \leq \varepsilon + \xi_m, \\ y_m - w^T \varphi(x_m) - b \geq \varepsilon - \xi_m^*, \\ \xi_m, \xi_m^* \geq 0, m = 1,\ldots, n, \end{cases} \quad (7)$$

where $C$ is the penalty parameter indicating the trade-off between the empirical risk and the regularization term.

The generic equation, using Lagrange multipliers and the optimality constraints is given as:

$$g(x) = \sum_{m=1}^{n} \left( \beta_m - \beta_m^* \right) K(x_m, x) + b \quad (8)$$

where the non-zero Lagrange multipliers and kernel function are $\beta_m$ and $\beta_m^*$, and $K(x_m, x)$ respectively. We tested a range of kernel functions in this study and the Radial Basis Function kernel (RBF) yield the best performance.

$$K(x_m, x_i) = \exp\left( -\gamma ||x_m - x_i||^2 \right), \quad (9)$$

where $\gamma$ stand for the width parameter of RBF.

**Table 1.** Experimental dataset [26]

| Mixtures | Coarse aggregate size (mm) | Cement (kg) | W/C (kg/m3) | Coarse aggregate (kg) | Density (kg/m3) | Compressive strength (MPa) | Tensile strength (MPa) | Porosity (%) |
|---|---|---|---|---|---|---|---|---|
| C1 | 4.5 | 200 | 0.35 | 1600 | 1780.31 | 2.21 | 0.36 | 35 |
| C2 | 9.5 | 200 | 0.35 | 1600 | 1697.50 | 2.45 | 0.36 | 38 |
| C3 | 12.5 | 200 | 0.35 | 1600 | 1723.38 | 3.22 | 0.65 | 38 |
| C4 | 14 | 200 | 0.35 | 1600 | 1637.62 | 2.42 | 0.32 | 42 |
| C5 | 17 | 200 | 0.35 | 1600 | 1716.13 | 3.53 | 0.47 | 35 |
| C6 | 22 | 200 | 0.35 | 1600 | 1683.46 | 2.8 | 0.44 | 38 |
| C7 | 9.5 | 200 | 0.3 | 1600 | 1682.72 | 1.79 | 0.21 | 39 |
| C8 | 9.5 | 200 | 0.4 | 1600 | 1752.95 | 3.92 | 0.79 | 35 |
| C9 | 12.5 | 200 | 0.3 | 1600 | 1652.40 | 2.43 | 0.34 | 39 |
| C10 | 12.5 | 200 | 0.4 | 1600 | 1767.00 | 3.67 | 0.84 | 36 |
| C11 | 22 | 200 | 0.3 | 1600 | 1662.02 | 1.91 | 0.48 | 39 |
| C12 | 22 | 200 | 0.4 | 1600 | 1796.57 | 3.59 | 0.75 | 35 |
| C13 | 9.5 | 150 | 0.35 | 1800 | 1679.76 | 1.16 | 0.2 | 40 |
| C14 | 12.5 | 150 | 0.35 | 1800 | 1652.40 | 1.06 | 0.21 | 40 |
| C15 | 22 | 150 | 0.35 | 1800 | 1665.71 | 1.83 | 0.46 | 38 |
| C16 | 9.5 | 250 | 0.35 | 1800 | 1809.14 | 4.76 | 0.66 | 35 |
| C17 | 12.5 | 250 | 0.35 | 1800 | 1708.59 | 3.07 | 0.65 | 37 |
| C18 | 22 | 250 | 0.35 | 1800 | 1715.25 | 3.45 | 0.69 | 36 |
| C19 | 9.5 | 150 | 0.35 | 1500 | 1679.76 | 1.71 | 0.32 | 38 |
| C20 | 12.5 | 150 | 0.35 | 1500 | 1656.10 | 1.67 | 0.48 | 38 |
| C21 | 22 | 150 | 0.35 | 1500 | 1678.28 | 1.71 | 0.56 | 38 |
| C22 | 9.5 | 250 | 0.35 | 1500 | 1804.71 | 6.95 | 1.32 | 30 |
| C23 | 12.5 | 250 | 0.35 | 1500 | 1707.85 | 5.14 | 0.81 | 32 |
| C24 | 22 | 250 | 0.35 | 1500 | 1874.94 | 6.45 | 1.2 | 31 |

## 3. Methodology
### 3.1 Description of dataset

Experimental results obtained by Ibrahim et al. [26] served as the datasets applied in this study. The experimental data has twenty-four data instances of PC mixtures, and the details are shown in Table 1. The descriptive input variables include the coarse aggregate size, cement mass, water/cement ratio and coarse aggregate mass, while the prediction outputs are the density, compressive strength, tensile strength and porosity of the PC. Thus, four (4) different XGBoost models are developed to train and predict the PC density, compressive strength, tensile strength and porosity respectively. The basic statistics of the dataset in terms of the mean, standard deviation, minimum value, lower quartile, middle quartile (median), upper quartile, and maximum value of each variable is presented in Table 2 to show its consistency and suitability for use in this study.

**Table 2.** Statistical Analysis Result [26]

|  | Coarse aggregate size (mm) | Cement (kg) | W/C (kg/m3) | Coarse aggregate (kg) | Density (kg/m3) | Compressive strength (MPa) | Tensile strength (MPa) | Porosity (%) |
|---|---|---|---|---|---|---|---|---|
| Count | 24 | 24 | 24 | 24 | 24 | 24 | 24 | 24 |
| Mean | 14.313 | 200 | 0.35 | 1625 | 1716.023 | 3.0375 | 0.565 | 36.75 |
| Standard deviation | 5.526 | 36.116 | 0.026 | 111.316 | 61.29 | 1.55 | 0.289 | 2.893 |
| Minimum value | 4.5 | 150 | 0.3 | 1500 | 1637.62 | 1.06 | 0.2 | 30 |
| 25% | 9.5 | 187.5 | 0.35 | 1575 | 1675.138 | 1.82 | 0.355 | 35 |
| 50% | 12.5 | 200 | 0.35 | 1600 | 1702.675 | 2.625 | 0.48 | 38 |
| 75% | 22 | 212.5 | 0.35 | 1650 | 1756.463 | 3.61 | 0.705 | 38.25 |
| Maximum value | 22 | 250 | 0.4 | 1800 | 1874.94 | 6.95 | 1.32 | 42 |

### 3.3. Evaluation Metrics

To evaluate the performance of the developed machine learning models in this study, we applied widely accepted metrics such as the square of correlation coefficient ($R^2$), root mean squared error (RMSE), mean absolute error (MAE) and mean absolute percentage error (MAPE). The mathematical formulation for each of these metrics is presented in Equations (10) to (13).

$$R^2 = \left( \frac{\frac{1}{n}\sum_{m=1}^{n}(\hat{y}_m - \bar{\hat{y}})(y_m - \bar{y})}{\sqrt{\left(\frac{1}{n}\sum_{m=1}^{n}(\hat{y}_m - \bar{\hat{y}})^2\right)\left(\frac{1}{n}\sum_{m=1}^{n}(y_m - \bar{y})^2\right)}} \right)^2 \quad (10)$$

$$\text{RMSE} = \sqrt{\frac{1}{n}\sum_{m=1}^{n}(y_m - \hat{y}_m)^2} \quad (11)$$

$$\text{MAE} = \frac{1}{n}\sum_{m=1}^{n}|y_m - \hat{y}_m| \quad (12)$$

$$\text{MAPE} = \frac{1}{n}\sum_{m=1}^{n}\frac{|y_m - \hat{y}_m|}{max(\varepsilon, |y_m|)} \quad (13)$$

where $y_m$ is the experimental output, $\hat{y}_m$ is the model estimated output, $\bar{y}$ is the mean of the experimental output, $\bar{\hat{y}}$ is the mean of the estimated output, $n$ is the number of samples. MAPE also has $\varepsilon$ which stands for an arbitrarily positive small constant to avoid division by zero when $y_m$ is zero. For each of MAPE, RMSE and MAE, the lower the value, the better model. On the contrary, the closer the $R^2$ value to 1 the better.

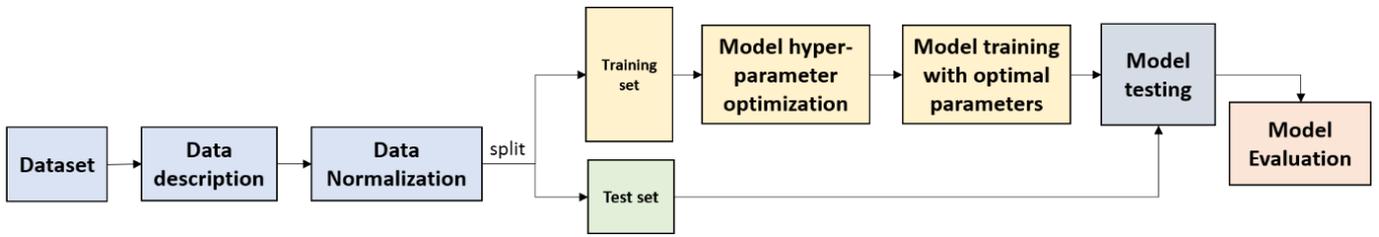

Fig. 1. Research Flow

### 3.2. Experimental design

The overall experimental design of this study is illustrated in Fig. 1, where data description that entails exploratory analysis of the dataset via its basic statistics (Table 2) is shown to precede data normalization. Both the input features and their respective output were normalized to a range of 0 and 1 to ensure a uniform range across features. Similar to Adewumi et al. [13], a training-test split of 80-20% is employed in this study; using the same training and test set to train and evaluate the performance of both the XGBoost and SVR models respectively. Hyperparameter optimization of each model is also carried out to ensure optimal model performance. It should be noted that only the training set is used in hyperparameter optimization. Contrary to Adewumi et al. [13] and Al-Sodani et al. [27] where the set of hyperparameters that yield the best performance on the test set was chosen, we adopt a rather more objective approach where the test set is not involved in model optimization. We adopt this to mimic the real-life scenario where the actual outputs of input data are usually unknown.

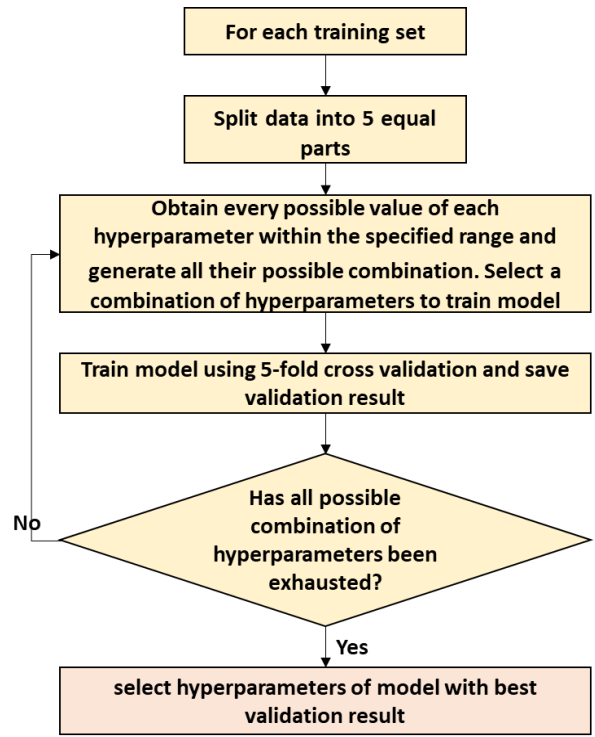

Fig. 2. Model Hyperparameter Optimization

Fig. 2 shows the procedure adopted for hyperparameter optimization in this study. A grid search approach, which is essentially an exhaustive search of every possible combination of the values of selected hyperparameters to identify the best combination that yields the most optimal model performance. A 5-fold cross validation is used to evaluate the performance of every combination of hyperparameters to identify the one that produced the best mean squared error over the 5-folds. The optimal parameters over this process are used to train the final model. For each of XGBoost and SVR machine learning algorithm, four models are optimized for the prediction of Density ($kg/m^3$), Compressive strength (MPa), Tensile strength (MPa) and Porosity (%) respectively.

For the SVR models, using the python scikit-learn machine learning programming library [28], four hyperparameters that include the regularization parameter (C), gamma (ɣ), epsilon (ε) and kernel function were optimized over a range of 1-200 at interval 1, 0 to 1 at interval 0.00001, 0 to 1 at interval 0.00001 and linear, polynomial, radial basis function and sigmoid respectively. On the other hand, XGBoost has a wider range of tuneable hyperparameters from which some were selected for optimization in this study. Leveraging the scikit-learn wrapper application programming interface for XGBoost, the following hyperparameters were exhaustively searched via grid search for the XGBoost models; maximum tree depth for base learners (max_depth), number of gradient boosted trees (n_estimators), subsample ratio of columns when constructing each tree (colsample_bytree), minimum loss reduction required to make a further partition on a leaf node of the tree (gamma), L1 regularization term on weights (reg_alpha), L2 regularization term on weights (reg_lambda) and learning rate (eta). The optimal set of parameters for each model are presented in Table 3. For notational convenience, the optimal XGBoost and SVR models for estimating Density (kg/m$^3$), Compressive strength (MPa), Tensile strength (MPa), Porosity (%) are respectively denoted as XGBoost1, XGBoost2, XGBoost3, XGBoost4 and SVR1, SVR2, SVR3, SVR4 models respectively.

**Table 3.** Optimal Parameters for the Developed Models

|  | Density (kg/m$^3$) | Compressive strength (MPa) | Tensile strength (MPa) | Porosity (%) |
|---|---|---|---|---|
| **XGBoost** | **XGBoost1** | **XGBoost2** | **XGBoost3** | **XGBoost4** |
| n_estimators | 100 | 18 | 25 | 83 |
| colsample_bytree | 1 | 0.7 | 0.7 | 0.7 |
| max_depth | 5 | 5 | 5 | 5 |
| subsample | 0.7 | 1 | 1 | 1 |
| reg_alpha | 1.1 | 0.11 | 0.11 | 0.02 |
| reg_lambda | 1.65 | 1.69 | 0.81 | 0.92 |
| gamma | 0.005 | 0.001 | 0.01 | 0.002 |
| eta | 0.3 | 0.95 | 0.34 | 0.28 |
| **SVR** | **SVR1** | **SVR2** | **SVR3** | **SVR4** |
| $C$ | 1 | 3 | 39 | 29 |
| gamma | 0.16687 | 0.02 | 0.11 | 0.117 |
| epsilon | 0.24 | 0.1 | 0.1 | 0.01 |
| kernel | rbf | rbf | rbf | rbf |

## 4. Results

XGBoost machine learning algorithm is used for the task of estimating the density (kg/m$^3$), compressive strength (MPa), tensile strength (MPa) and porosity (%) of the different PC mixtures in the dataset. This is accomplished by using 4 models altogether. The performance of each model in terms of the afore-detailed evaluation metrics are presented in Table 4 for the training and test phases. PC Mixtures C11, C12, C15, C21 and C23 were used to test models while the rest were used for training. The performance of each model on each estimation task is further elaborated in the following subsections. In addition, four SVR models are also developed as a baseline for comparing the results from the four XGBoost models. This comparison is to evaluate the performance of XGBoost in estimating the properties of PC. A sensitivity analysis of the variables is presented in what follows as a precursor to detailed explanation of the model estimations for each prediction task.

### 4.1. Sensitivity Analysis

A sensitivity analysis of the XGboost and SVR models to all the input variables employed in estimating the input-output relationship is presented here. This is necessary to understand how changes in the input variable bring about corresponding changes in output. Thus, similar to [18], the sensitivity each mixture of PC (i.e. Nominal Coarse aggregate size, Cement, W/C and Coarse aggregate mm)) to Density (kg/m$^3$), Compressive strength (MPa), Tensile Strength (MPa) and Porosity (%) is respectively assessed using correlation coefficient (CC) which is essentially the square root of Equation 10. In relation to Density (kg/m$^3$) (Fig. 3(a)), each of the input variables have some degree of correlation with the output given that two input variables (Cement and W/C) have positive CC (i.e., direct correlation) whereas the other two (Nominal Coarse aggregate size (mm) and Coarse aggregate (kg)) are inversely correlated to the same output. The implication of this is that any increase in the former results in corresponding increase in Density whereas an increase in latter decreases the Density. Similar pattern can be observed with respect to Compressive strength (MPa) (Fig. 3 (b)) and Tensile Strength (MPa) (see Fig. 3(c)) as outputs except that, in addition to the positively correlated inputs (Cement and W/C variables), nominal Coarse aggregate size also has positive CC with the respective outputs; indicating positive correlation. However, in relation to output variable (Fig. 3 (d)), only

Coarse aggregate (kg) is positively correlated with the output while the three remaining input variables (Nominal Coarse aggregate size (mm), Cement and W/C variables) are inversely correlated to it. Generally, the input variables are correlated with the respective output, however the Nominal Coarse aggregate size (mm) input appear to have the least or minimal correlation; particularly in relation to Porosity (%) (as in Fig. 3(d)). In this study, we do not assume independence between input variables, given that strong correlation can exist between them. Thus, an inclusion of all the input variables (Nominal Coarse aggregate size (mm), Cement, W/C variables and Coarse aggregate (kg)) is suggested in this study for improved estimation of the respective outputs by the models.

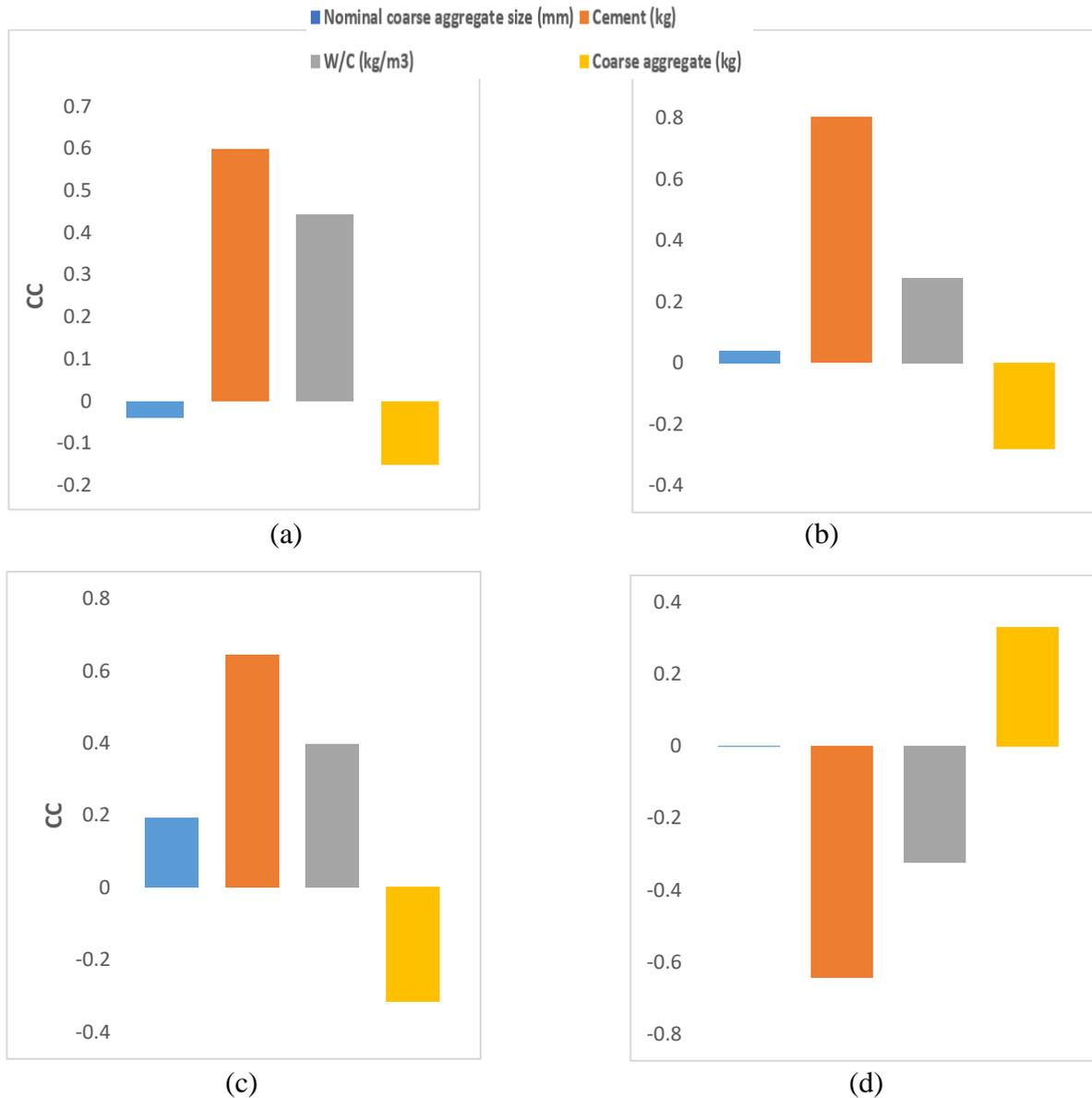

Fig. 3. Correlation of Input variables with (a) Density (kg/m$^3$), (b) Compressive strength (MPa), (c) Tensile Strength (MPa) and (d) Porosity (%).

### 4.1. Estimation of Density (kg/m$^3$)

From Table 4, the values of the evaluation metrics for density estimation using XGBoost1 and SVR1 models are shown. In the training and test phases of XGBoost1, the $R^2$ are 0.7332 and 0.5994, RMSE are 33.2279 kg/m$^3$ and 34.9663 kg/m$^3$, MAE are 24.5669 kg/m$^3$ and 27.8676 kg/m$^3$, and MAPE are 0.0141 and 0.0162 respectively. The low error values (RMSE, MAE and MAPE) at both training and test phases are indicative of the capability of the XGBoost1 model as an accurate estimator of density of PC.

**Table 4.** Results of Developed Models

| | Training | | | | Test | | | |
|---|---|---|---|---|---|---|---|---|
| **Density (kg/m³)** | | | | | | | | |
| | R² | RMSE (kg/m3) | MAE (kg/m3) | MAPE | R² | RMSE (kg/m3) | MAE (kg/m3) | MAPE |
| XGBoost1 | 0.7332 | **33.1294** | 24.9044 | 0.0143 | **0.5994** | **34.9663** | **27.8676** | **0.0162** |
| SVR1 | **0.8277** | 33.2279 | **24.5669** | **0.0141** | 0.3367 | 44.0579 | 32.8641 | 0.0190 |
| **Compressive strength (MPa)** | | | | | | | | |
| | R² | RMSE (MPa) | MAE (MPa) | MAPE | R² | RMSE (MPa) | MAE (MPa) | MAPE |
| XGBoost2 | **0.9787** | **0.2391** | **0.2117** | **0.0768** | 0.8973 | **0.5777** | **0.5220** | **0.2365** |
| SVR2 | 0.9288 | 0.4416 | 0.3178 | 0.1004 | **0.8993** | 0.7430 | 0.6494 | 0.2666 |
| **Tensile strength (MPa)** | | | | | | | | |
| | R² | RMSE (MPa) | MAE (MPa) | MAPE | R² | RMSE (MPa) | MAE (MPa) | MAPE |
| XGBoost3 | **0.9581** | **0.0695** | **0.0532** | **0.1193** | 0.8735 | **0.1727** | **0.1188** | **0.1831** |
| SVR3 | 0.9466 | 0.0738 | 0.0575 | 0.1405 | **0.8927** | 0.2127 | 0.1493 | 0.2308 |
| **Porosity (%)** | | | | | | | | |
| | R² | RMSE (%) | MAE (%) | MAPE | R² | RMSE (%) | MAE (%) | MAPE |
| XGBoost4 | **0.9761** | **0.5427** | **0.4776** | **0.0131** | **0.8622** | **0.9777** | **0.8956** | **0.0240** |
| SVR4 | 0.8180 | 1.2346 | 0.6320 | 0.0167 | 0.8559 | 1.2764 | 1.1146 | 0.0313 |

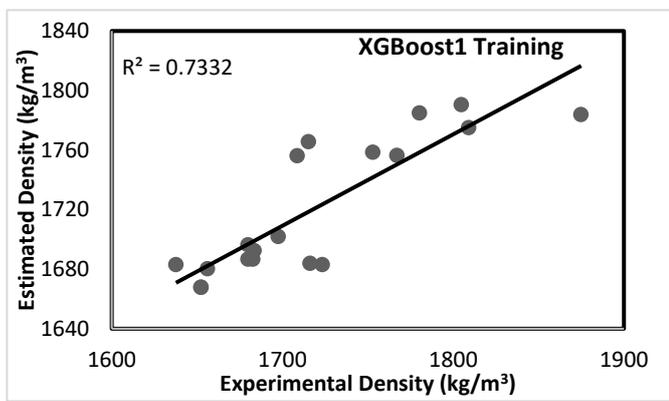

(a)

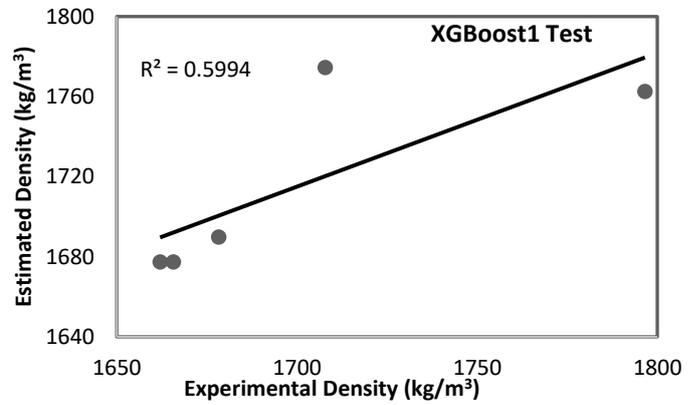

(b)

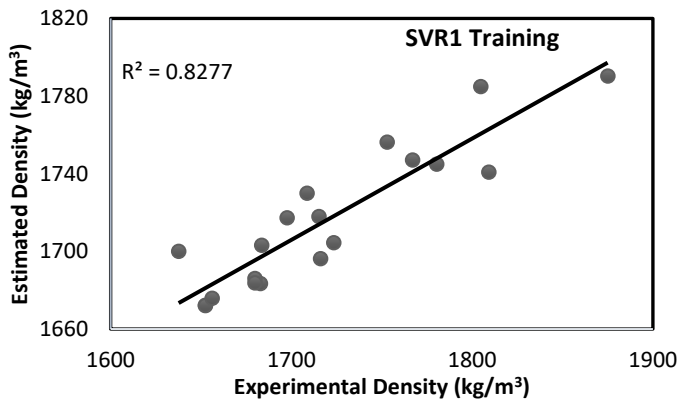

(c)

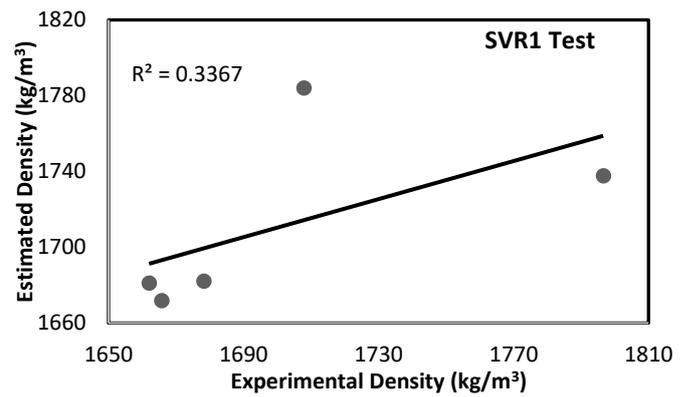

(d)

Fig. 4. Scatter Plots of Experimental against Estimated Density (kg/m³)

Comparing the evaluation metric values obtained for XGBoost1 to SVR1 ($R^2$ = 0.8277, RMSE = 33.1294 kg/m³, MAE = 24.9044 kg/m³, MAPE = 0.0143), both models show comparable performance in the training phase with SVR1 marginally performing better than XGBoost1 in three of the four evaluation metrics. However, in the test phase, XGBoost1 model ($R^2$ = 0.5994, RMSE = 34.9663 kg/m³, MAE = 27.8676 kg/m³, MAPE = 0.0162) outperforms SVR1 ($R^2$ = 0.3367, RMSE = 44.0579 kg/m³, MAE = 32.8641 kg/m³, MAPE = 0.0190) across all the evaluation metrics with performance improvement of 14%, 15%, and 20% in terms of MAPE, MAE and RMSE respectively. In comparison to the test results from [13] where the same data had been utilized, XGBoost1 also shows better test performance in terms of RMSE and MAE while showing lower $R^2$.

Fig. 4 presents the scatter plots of the estimated against experimental densities with their respective regression lines for the training and test phases of XGBoost1 and SVR1 models. This shows how correlated the estimated values are to the experimental ones. The $R^2$ which is given by the square of correlation coefficient as in Equation (10) summarises this relationship in a single value. Although, XGBoost1 density estimations show lesser correlation compared to SVR1 at training phase, the former produces density estimates that are more correlated to the experimental values in the test phase with as much as 78% improvement over the latter in terms of $R^2$.

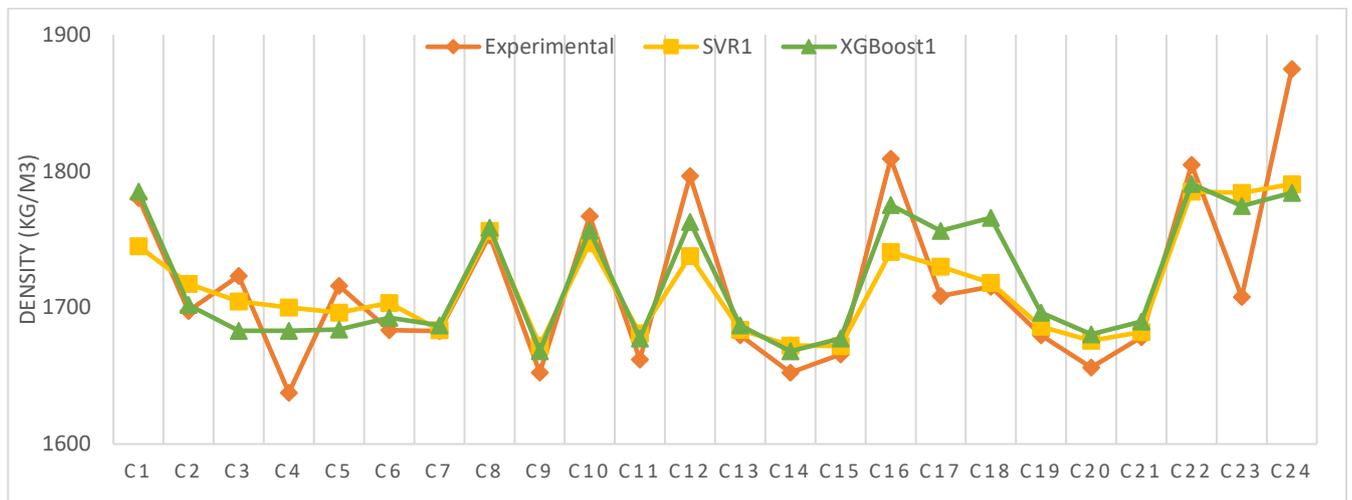

Fig. 5. Comparison of the Experimental and Estimated Density (kg/m³)

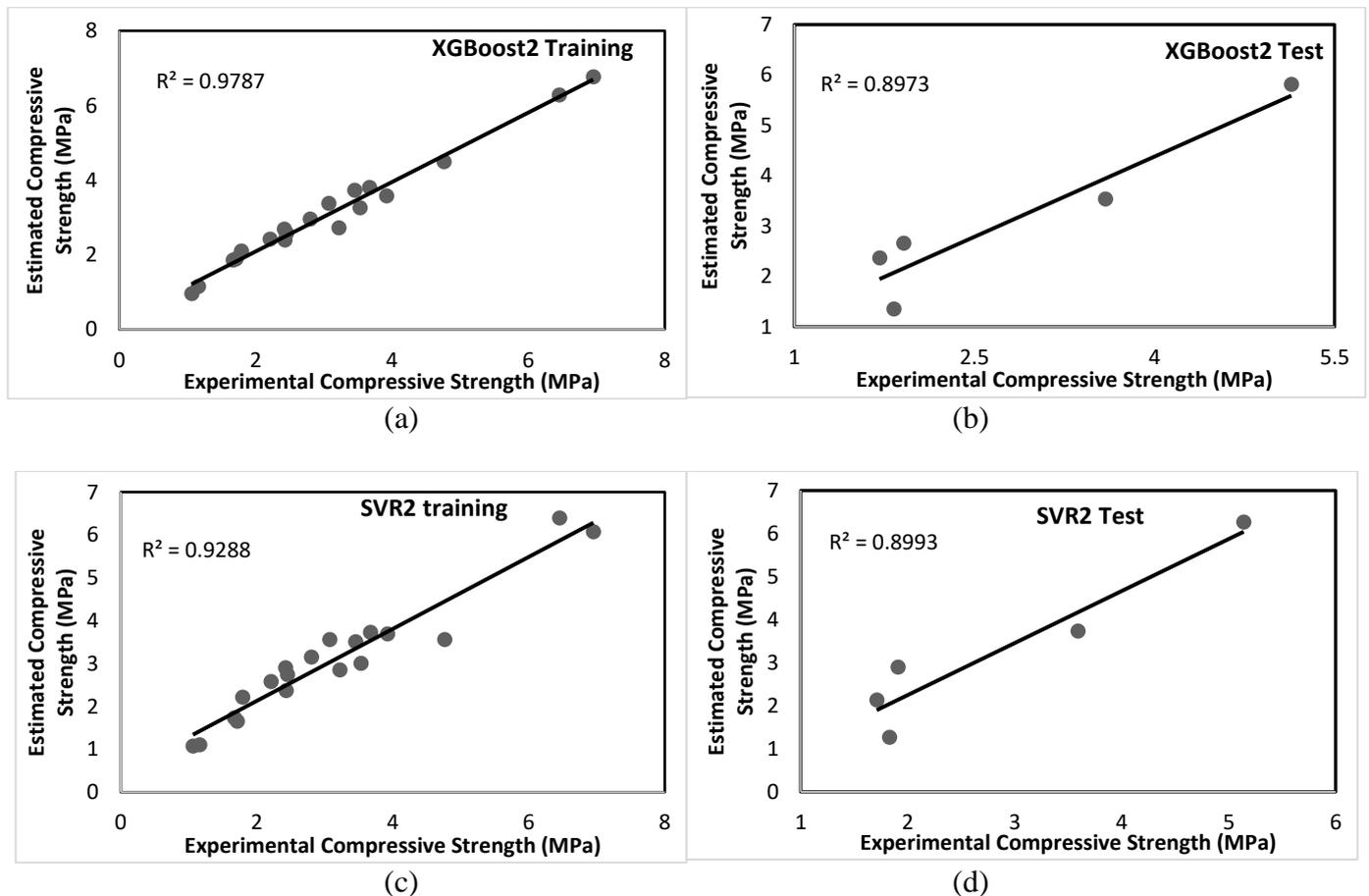

Fig. 6. Scatter Plots of Experimental against Estimated Compressive Strength (MPa)

Line plots of experimental, SVR1-estimated and XGBoost1-estimated densities for the various PC mixtures are presented in Fig. 5. While both models (XGBoost1 and SVR1) generally produced impressive and similar estimates to the experimental density, XGBoost1 produced more accurate results. For example, the experimental density value of mixture 12 (C12) is 1796.57 kg/m$^3$, while the XGBoost1 and SVR1 estimated values are 1762.71 kg/m$^3$ and 1737.70 kg/m$^3$ respectively. Similarly, the experimental density of C23 is 1707.85 kg/m$^3$, while the XGBoost1 and SVR1 estimated values are 1774.61 kg/m$^3$ and 1784/38 kg/m$^3$ respectively.

## 4.2. Estimation of Compressive Strength (MPa)

Table 4 provides the values of the evaluation metrics for estimation of the compressive strength of PC using XGBoost2 and SVR2 models. In the training and test phases of XGBoost2, the $R^2$ are 0.9787 and 0.8973, RMSE are 0.2391 MPa and 0.5777 MPa, MAE are 0.2117 MPa and 0.5220 MPa, and MAPE are 0.0768 and 0.2365 respectively. The low error values (RMSE, MAE and MAPE) at both training and test phases shows the potential of the XGBoost2 model in accurately predicting compressive strength of PC.

Comparing the evaluation metric values obtained for XGBoost2 to SVR2 ($R^2$ = 0.9288, RMSE = 0.4416 MPa, MAE = 0.3178 MPa, MAPE = 0.1004), XGBoost2 showed better performance than SVR2 across all evaluation metrics in the training phase. Similarly, the XGBoost2 ($R^2$ = 0.8973, RMSE = 0.5777 MPa, MAE = 0.5220 MPa, MAPE = 0.2365) produced better performance than the SVR2 ($R^2$ = 0.8993, RMSE = 0.7430 MPa, MAE = 0.6494 MPa, MAPE = 0.2666) in the test phase across all but for $R^2$ metric where SVR2 marginally performed better with 0.2% improvement. Compared to results from [13, 26], the XGBoost2 model produced superior performance across the evaluation metrics.

A scatter plot of the experimental and estimated compressive strengths of both models for the training and test phases is presented in Fig. 6. The potential of both models to estimate the compressive strength of the various PC mixtures with high level of accuracy is further shown by the high $R^2$ values.

In Fig. 7, a line plot of the experimental, XGBoost2-estimated and SVR2-estimated compressive strengths (MPa) shows that both models generally produced impressive and similar estimates to the experimental compressive strength with XGBoost2 proving better of the two models. For example, the experimental compressive strength of PC mixture 11 (C11) is 1.91 MPa, while the XGBoost2 and SVR2 estimated values are 2.65 MPa and 2.90 MPa respectively. Similarly, the experimental compressive strength of C23 is 5.14 MPa, while the XGBoost2 and SVR2 estimated values are 5.81 MPa and 6.26 MPa respectively.

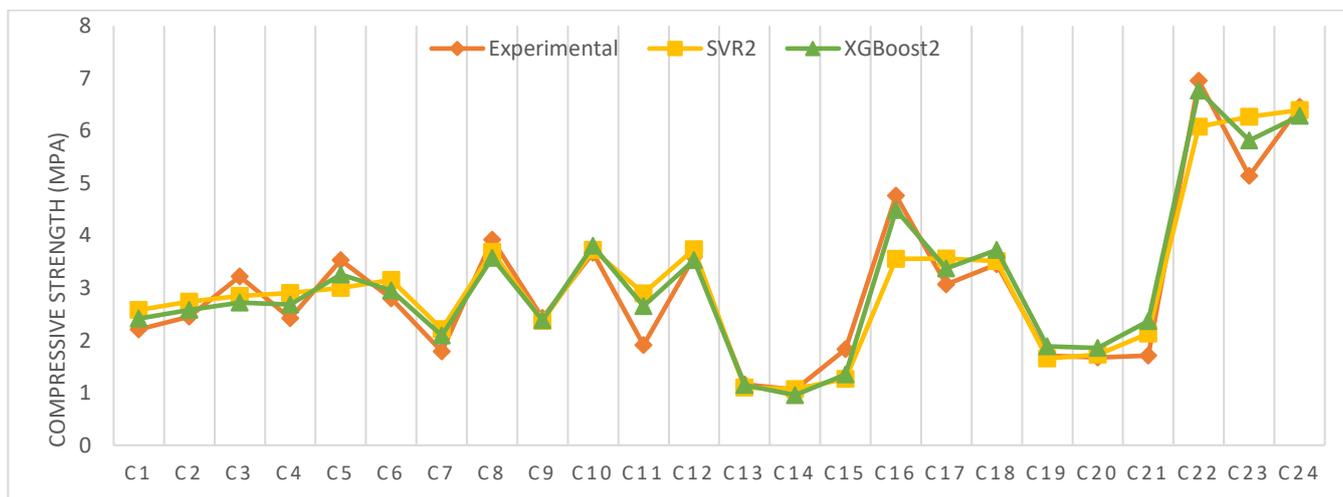

Fig. 7. Comparison of the Experimental and Estimated Compressive Strength (MPa)

## 4.3. Estimation of Tensile Strength (MPa)

In Table 4, the values of the evaluation metrics for estimation of the tensile strength of PC using XGBoost3 and SVR3 models are shown. In the training and test phases of XGBoost3, the $R^2$ are 0.9581 and 0.8735, RMSE are 0.0695 MPa and 0.1727 MPa, MAE are 0.0532 MPa and 0.1188 MPa, and MAPE are 0.1193 and 0.1831 respectively. The low error values (RMSE, MAE and MAPE) at both training and test phases show the capability of the XGBoost3 model in estimating the tensile strength of PC with high accuracy.

Comparing the evaluation metric values obtained for XGBoost3 ($R^2$ = 0.9581, RMSE = 0.0695 MPa, MAE = 0.0532 MPa, MAPE = 0.1193) to SVR3 ($R^2$ = 0.9466, RMSE = 0.0738 MPa, MAE = 0.0575 MPa, MAPE

= 0.1405), XGBoost3 showed slightly superior performance than SVR3 in the training phase. However, in the test phase, the XGBoost3 model ($R^2$ = 0.8735, RMSE = 0.1727 MPa, MAE = 0.1188 MPa, MAPE = 0.1831) significantly outperforms its SVR3 ($R^2$ = 0.8927, RMSE = 0.2127 MPa, MAE = 0.1493 MPa, MAPE = 0.2308) counterpart across all metric with at 18% improvement except in terms of $R^2$ where the SVR3 model showed about 2% improvement. Furthermore, the scatter plots presented in Fig. 8 shows models have the capability of estimating tensile strength with impressive accuracy given the high degree of correlation shown in the training and test phases. The high $R^2$ score shows better SVR3 performance in this regard. In comparison to what has been reported in [13], the performance of both the XGBoost3 and SVR3 models across all but $R^2$ is slightly lower.

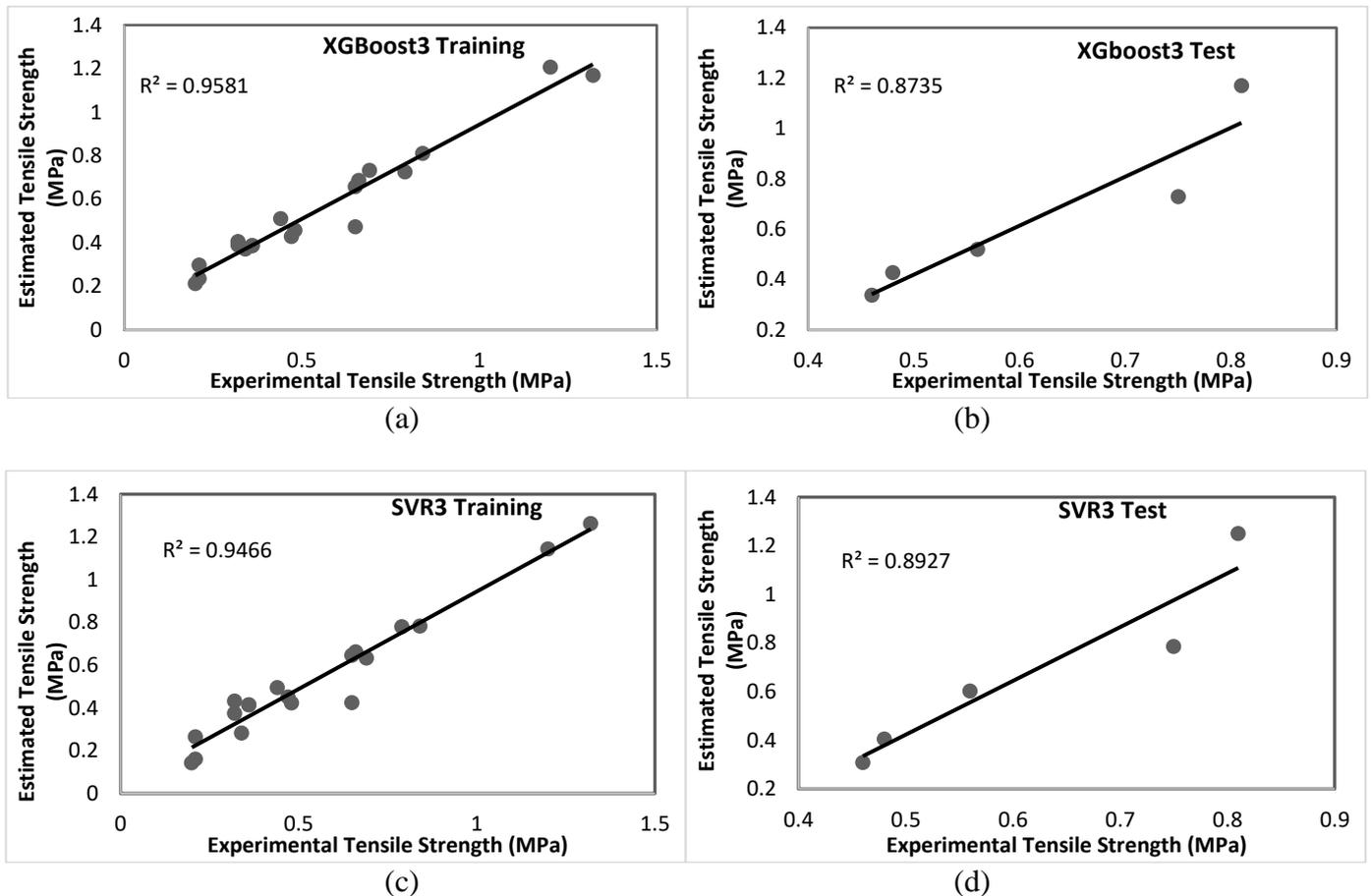

Fig. 8. Scatter Plots of Experimental against Estimated Tensile Strength (MPa)

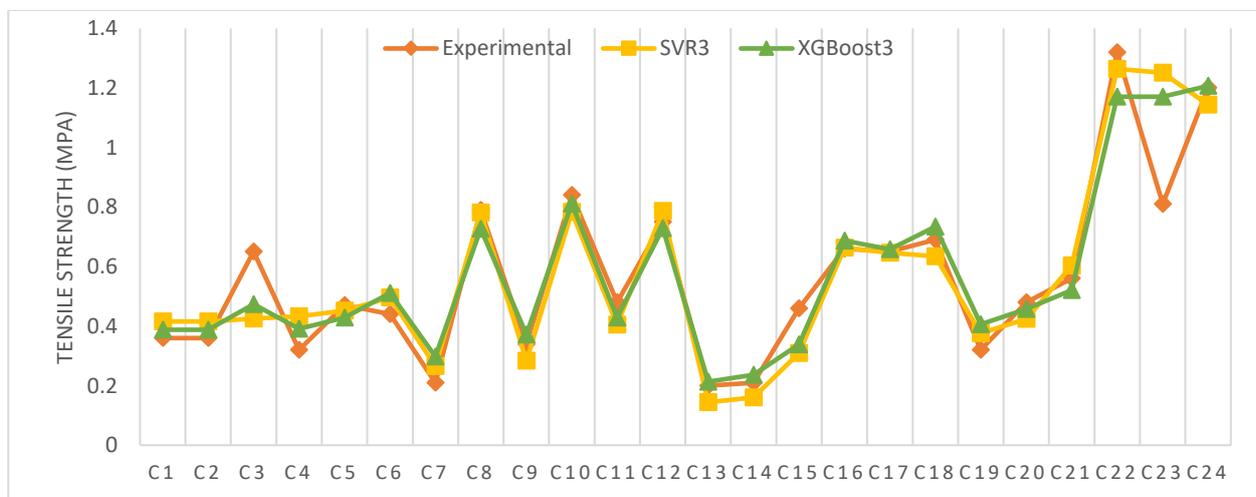

Fig. 9. Comparison of the Experimental and Estimated Tensile Strength (MPa)

In addition, the line plot presented in Fig. 9 also attest to the impressive performance of both models (XGBoost3 and SVR3) with XGBoost3 producing more accurate results. For example, the experimental tensile strength of mixture 11 (C11) is 0.48 MPa, while the XGBoost3 and SVR3 estimated values are 0.43

MPa and 0.41 MPa respectively. Similarly, the experimental tensile strength of C23 is 0.81 MPa, while the XGBoost3 and SVR3 estimated values are 1.17 MPa and 1.25 MPa respectively.

## 4.4. Estimation of Porosity (%)

Table 4 also shows the values of the evaluation metrics for estimation of the porosity of PC using XGBoost4 and SVR4. In the training and test phases of XGBoost4, the $R^2$ are 0.9761 and 0.8622, RMSE are 0.5427 % and 0.9777 %, MAE are 0.4776 % and 0.8956 %, and MAPE are 0.0131 and 0.0240 respectively. The low error values (RMSE, MAE and MAPE) at both training and test phases are indicative of the potentials of the XGBoost4 in accurately predicting the tensile strength of the PC.

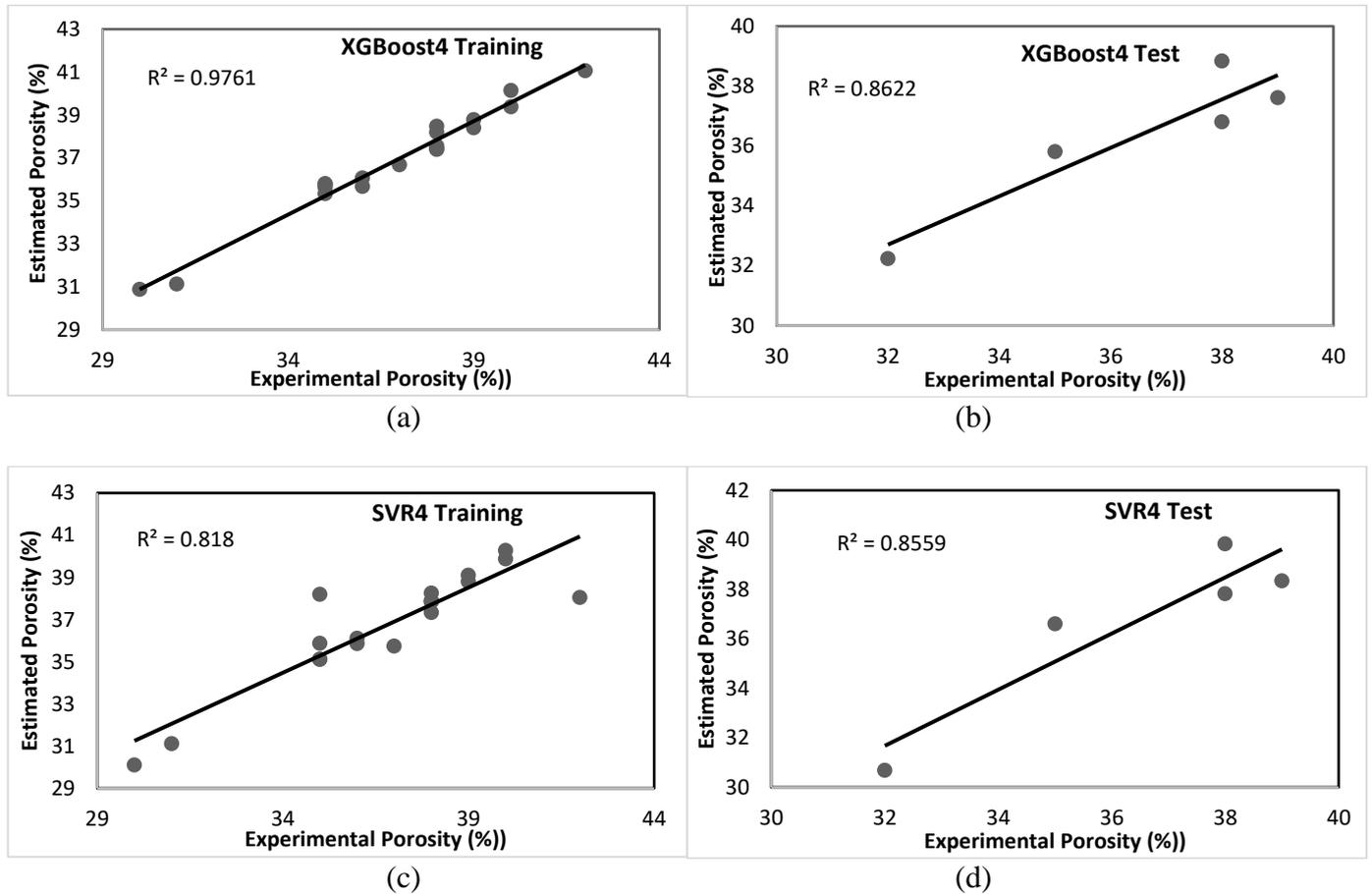

Fig. 10. Scatter Plots of Experimental against Estimated Porosity (%)

Comparing the evaluation metric values obtained for XGBoost4 ($R^2$ = 0.9761, RMSE = 0.5427 %, MAE = 0.4776 %, MAPE = 0.0131) to SVR4 ($R^2$ = 0.8180, RMSE = 1.2346 %, MAE = 0.6320 %, MAPE = 0.0167), XGBoost4 performed better than the SVR4 model in the training phase. Similar performance is also shown by the XGBoost4 model ($R^2$ = 0.8622, RMSE = 0.9777 %, MAE = 0.8956 %, MAPE = 0.0240) over the SVR4 model ($R^2$ = 0.8559, RMSE = 1.2764 %, MAE = 1.1146 %, MAPE = 0.0313) in the test phase; achieving performance improvement that ranges between 19% and 23% in terms of RMSE, MAE and MAPE. In the scatter plot presented in Fig. 10, the correlation of the estimated porosity with the experimental values is illustrated for the training and test phases of both models. Although, the $R^2$ value of the SVR4 model is slightly lower than the XGBoost4 model, the performance of both models is impressive; exceeding what has been reported on the same dataset in pertinent works [13].

In addition, Fig. 11 presents a line plot of the experimental, XGBoost4-estimated and SVR4-estimated porosity values. While the estimation of both models are impressively close to the corresponding experimental values, the XGBoost4 model produced a better estimate of the actual values. For example, the experimental porosity of mixture 12 (C12) is 35%, while the XGBoost4 and SVR4 estimated values are 35.81% and 36.61 respectively. Similarly, the experimental porosity of C23 is 32%, while the XGBoost4 and SVR4 estimated values are 32.24% and 30.70% respectively.

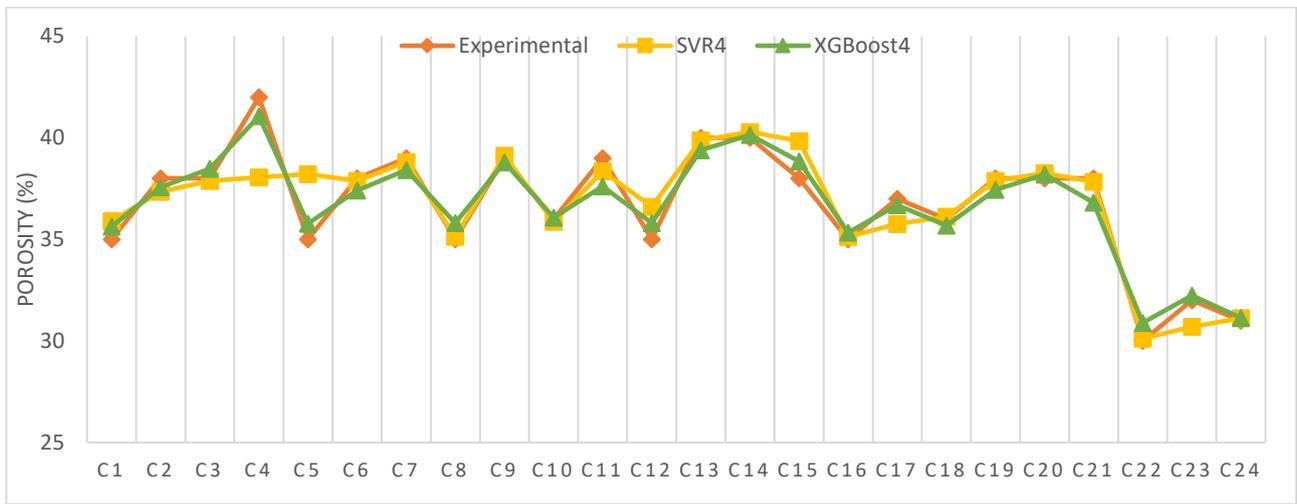

Fig. 11. Comparison of the Experimental and Estimated Porosity (%)

In all cases, the computational models generally performed impressively with the training results characteristically better than the corresponding test results except in the case of $R^2$ of the SVR4 model. Nonetheless, the training performance of the models are expected to be better than the corresponding test performance because the models were exposed to the actual outputs during the training phase. This is why a nonbiased measure of the performance of the models are the test results, and as shown from the foregoing, both the SVR and XGBoost-based models show impressive estimates of the experimental values with the XGBoost models showing better estimates overall. Moreover, a comparison of the XGBoost-estimated properties with the what has been reported in pertinent works [13, 26] reveal better performance.

**Table 5.** Feature Importance Ranking

| Model | Feature | Rank | | | Mean rank |
|---|---|---|---|---|---|
| | | Gain | Weight | Cover | |
| XBGoost1 | | | | | |
| | Cement (kg) | 1 | 3 | 1 | 1.67 |
| | The nominal coarse aggregate size (mm) | 2 | 2 | 3 | 2.33 |
| | W/C (kg/m$^3$) | 3 | 1 | 4 | 2.67 |
| | Coarse aggregate (kg) | 4 | 4 | 2 | 3.33 |
| XBGoost2 | | | | | |
| | Cement (kg) | 1 | 3 | 1 | 1.67 |
| | The nominal coarse aggregate size (mm) | 4 | 1 | 3 | 2.67 |
| | W/C (kg/m$^3$) | 3 | 4 | 2 | 3.00 |
| | Coarse aggregate (kg) | 2 | 2 | 4 | 2.67 |
| XBGoost3 | | | | | |
| | Cement (kg) | 1 | 1 | 2 | 1.33 |
| | The nominal coarse aggregate size (mm) | 4 | 2 | 4 | 3.33 |
| | W/C (kg/m$^3$) | 3 | 4 | 1 | 2.67 |
| | Coarse aggregate (kg) | 2 | 3 | 3 | 2.67 |
| XBGoost4 | | | | | |
| | Cement (kg) | 1 | 2 | 1 | 1.33 |
| | The nominal coarse aggregate size (mm) | 4 | 1 | 4 | 3.00 |
| | W/C (kg/m$^3$) | 3 | 4 | 3 | 3.33 |
| | Coarse aggregate (kg) | 2 | 3 | 2 | 2.33 |

## 5. Feature Importance/Contribution with XGBoost

Model interpretation is crucial for understanding, validating, and interpreting machine-learning models [29]; one of the major advantages of XGBoost over SVR. Although, earlier works on machine learning-based prediction of properties of PC have rarely attempted analysis of the predictor variables contributions, this study sets the precedent for an interesting area of research. Being an ensemble of CART, XGBoost implementations provide a medium for researchers to gain some degree of insight or understanding of what contributed to the model performance. XGBoost allows ranking of features based on their importance/contribution to the prediction task. Feature importance ranking in XGBoost are calculated based on either gain, weight or cover. For gain, features are ranked based on the average gain over all splits where the feature is used in the ensemble. Weight is calculated based on the number of times a feature appears in a tree while cover refers to the average number of samples affected by the splits that use a feature. Each of these feature ranking methods were used to rank the features of each of the four XGBoost models in this study. Given the variation in ranking that could result from the different feature importance calculation methods, the average ranking of each feature across the three methods is a provided in Table 5 alongside the basic methods. The lower the value of the rank, the better.

From Table 5, the feature with highest mean rank across all the four XGBoost models is cement proportion (kg). This indicates that the quantity of cement is a crucial component in the various mixtures. On the other hand, the feature with least mean rank varies across the four XGBoost models with coarse aggregate proportion (kg), W/C (kg/m$^3$), nominal coarse aggregate size (mm), and W/C (kg/m$^3$) having the least average ranking for the XBGoost1, XBGoost2, XBGoost3 and XBGoost4 models respectively.

# 7. Conclusions

An advanced computational technique was applied to predict the density, compressive strength, tensile strength and porosity of pervious concrete (PC) by using XGBoost models 1, 2, 3 and 4 respectively. The results showed that the XGBoost models had low root mean square errors, low mean absolute errors and low mean absolute percentage errors, while the square of correlation coefficients were high. The developed XGBoost models predicted PC properties that were in good agreement with the experimental values. A comparison of the XGBoost predicted results (PC properties) with those from support vector regression (SVR) showed that predictions from XGBoost models were more accurate. The comparison indicated that XGBoost performed better than SVR with lower RMSE of 0.58 to SVR's 0.74 for compressive strength, 0.17 to SVR's 0.21 for tensile strength, 0.98 to SVR's 1.28 for porosity, and 34.97 to SVR's 44.06 for density. With the high correlation between the predicted and experimentally obtained properties, the XGBoost models are able to provide quick and reliable information on the properties of PC which are experimentally costly and time consuming. Furthermore, a feature importance and contribution analysis showed that cement proportion was the most significant factor in all the properties predicted. This study shows the potential of applying XGBoost to obtain accurate and quick estimation of PC properties in a very cost-effective way. A main limitation of this study is the small size of the dataset. Although, both models performed well despite the limited size, more data will likely bring performance improvement and this will be considered in future research endeavours.